\documentstyle[pre,aps,twocolumn]{revtex}

\begin{document}
\draft
\title{Distributions of Transition Matrix Elements in Classically Mixed
Quantum Systems}
\author{Dominique Boos\'e}
\address{Laboratoire de Physique Th\'eorique, Universit\'e Louis Pasteur,
F-67084 Strasbourg, France}
\author{J\"org Main}
\address{Institut f\"ur Theoretische Physik I, Ruhr-Universit\"at Bochum,
D-44780 Bochum, Germany}
%
\date{May 14, 1999}
\maketitle

\begin{abstract}
The quantitative contributions of a mixed phase-space to the mean 
characterizing the distribution of diagonal transition matrix elements and to 
the variance characterizing the distributions of non-diagonal transition 
matrix elements are studied. It is shown that the mean can be expressed as 
the sum of suitably weighted classical averages along an ergodic trajectory 
and along the stable periodic orbits. Similarly, it is shown that the values 
of the variance are well reproduced by the sum of the suitably weighted 
Fourier transforms of classical autocorrelation functions along an ergodic 
trajectory and along the stable periodic orbits. The illustrative numerical 
computations are done in the framework of the Hydrogen atom in a strong 
magnetic field, for three different values of the scaled energy.  
\end{abstract}


\section{Introduction}
It is well known that Random Matrix Theory is able to reproduce successfully 
statistical properties which characterize the semiclassical regime of quantum
systems with a few freedoms having a fully ergodic classical phase-space. 
This is true for bound systems since the short-ranged spectral statistics, 
those using energy levels which are separated by some multiples of the mean 
level spacing, are well described in this approach 
(see, e.g., Ref.\ \cite{Boh90}).
This is also true for scattering systems since the fluctuations of the 
$S$-matrix elements are well modeled with the help of random matrices 
\cite{Ver85}.
The predictions of Random Matrix Theory are universal, that is they are 
independent of the microscopic details of the particular system under study.
It is Classical Mechanics which provides the ultimate justification for 
universality. 
For instance, the universal behavior of the spectral statistics is justified 
by resorting to the principle that the very long unstable periodic orbits of 
the underlying classical dynamics are uniformly distributed in phase-space 
(see, e.g., Ref.\ \cite{Ber90}).
For instance also, the universal energy dependence of the $S$-matrix 
autocorrelation function is explained by the fact that the distribution in 
length of the very long unstable scattering trajectories has a generic 
functional form for classical ergodic systems (see, e.g., Ref.\ \cite{Smi90}).

Another example of universal behavior has to do with the transition matrix 
elements associated to an operator perturbing a quantum system with an 
ergodic phase-space. 
Indeed, Random Matrix Theory predicts that these transition matrix elements 
are independent random variables distributed in a normal way \cite{Por65}. 
However, mean and variance characterizing these distributions are free 
parameters and therefore lack a physical meaning in this approach. 
In view of the previously given examples, it appears natural to link this 
other universal property to Classical Mechanics again. 
This can actually be done by means of the generalization of the semiclassical 
theory pioneered by Gutzwiller \cite{Gut71,Gut90} to arbitrary (well behaved) 
operators \cite{Wil87,Eck92}, a generalization which allows to give a 
classical interpretation of mean and variance. 
Indeed, in the semiclassical limit of Quantum Mechanics, the leading order 
expression of the mean value of the diagonal transition matrix elements is 
equal to the microcanonical average of the Weyl transform of the perturbing 
operator \cite{Wil87,Eck92}.
In the same limit, the leading order expression of the variance associated 
to the Gaussian distributions generated with non-diagonal transition matrix 
elements is proportional to the Fourier transform of the classical 
autocorrelation function of the Weyl transform of the 
perturbing operator \cite{Wil87,Eck92}.
The link between classical variance and autocorrelation function has been 
studied numerically in three quantum systems with an ergodic phase-space, 
a quartic oscillator \cite{Aus92}, 
an oval billiard \cite{Meh95} and the Hydrogen atom in a strong magnetic  
field \cite{Boo95}.
In these systems and for several examples of perturbing operators, it has 
actually been found that the autocorrelation function of the 
Weyl transform of the perturbing operator determines with an excellent 
accuracy the local values of the classical variance.
Moreover, it has been verified \cite{Meh95} that the semiclassical 
corrections to the leading order expression of the variance which are due 
to the shortest unstable periodic orbits are in very good quantitative 
agreement with the theoretical predictions of Refs.\ \cite{Wil87,Eck92}.     
The autocorrelation function has also been studied in the
framework of a purely regular billiard for two examples of perturbing 
operators \cite{Min97}. It has to be added that a method has been proposed 
 \cite{Bog96} to compute the variance associated to the distribution of
diagonal transition matrix elements, a quantity which has not been considered 
in Refs.\ \cite{Wil87,Eck92}. This method, which rests on periodic orbit 
theory, has an interesting formal relationship with the supersymmetric methods
used in the description of disordered systems \cite{Alt95}. 

\vskip -1.5pt
Generic quantum systems with a small number of freedoms $d$ do have a mixed
regular-ergodic phase-space. This means that two different kinds of orbits
appear in the underlying classical dynamics.
There are orbits which wind regularly round $d$-dimensional tori and there 
are orbits which explore densely $(2d-1)$-dimensional regions of the energy 
shell in a highly chaotic manner.
Phase-space is therefore naturally partitioned into regular components, 
each including infinitely many neighboring tori, and ergodic components which 
are free of tori.
These various components are mutually independent.
Since Random Matrix Theory appears to be relevant to quantum systems with a 
fully ergodic phase-space exclusively, the previously listed universal 
properties do not hold anymore for quantum systems with a mixed phase-space.
However, one may still wonder about the presence in statistical properties 
pertaining to mixed quantum systems of fingerprints related to the coexistence
of regular and ergodic regions in phase-space.
Berry and Robnik \cite{Bro84} have been among the first to tackle this 
question. 
Assuming that the semiclassical spectrum of a mixed quantum system is the
superposition of statistically independent sequences of levels from each of 
the phase-space components, they have deduced a semi-phenomenological closed 
formula for the level spacing distribution. 
The formula makes use of the total phase-space volume of all regular regions 
as well as of the individual phase-space volumes of all ergodic regions. 
Their line of argument has been extended to other spectral 
statistics \cite{Sel85}.
The validity of the semi-phenomenological formulae obtained for the spectral 
statistics has been supported by numerical computations \cite{Sel85}. 
Apart from the one of Berry and Robnik \cite{Bro84}, two other 
parametrizations of the level spacing distribution have been proposed to
characterize a mixed quantum system. These are the
parametrizations of Brody \cite{Tbr73} and of Izrailev \cite{Izr87}, whose
validity has also been supported by numerical computations.   
The link between the mixed character of phase-space and statistical properties 
of transition matrix elements has been studied by Robnik and Prosen
\cite{Rob93,Pro94}.
Extending to eigenvectors the assumption of independent sequences originally 
proposed for energy levels, they have argued that the only transition matrix 
elements which have to be considered in the semiclassical limit are those 
whose initial and final states can be associated to the same component of 
phase-space. 
Moreover, they have generalized the results of Refs.\ \cite{Wil87,Eck92} 
by showing that the expression of the variance characterizing the 
distributions of non-diagonal transition matrix elements whose initial and 
final states can be associated to a given component of phase-space (which is 
either an ergodic region or some set of quantized tori lying closely together)
always involves a microcanonical average over this very component. 
Numerical computations \cite{Rob93,Pro94} have corroborated the validity of 
their results.

The purpose of the present paper is to provide a detailed study of the 
quantitative contributions coming from the various components of a mixed
phase-space to the mean characterizing the distribution of diagonal transition
matrix elements and to the variance characterizing the distributions of
non-diagonal transition matrix elements. It means in particular to complement 
the study of Robnik and Prosen on a specific aspect of the link between the 
mixed character of phase-space and statistical properties of transition matrix
elements which was not considered by them. 
This aspect concerns the use of the stable periodic orbits in the computation 
of the quantitative contributions coming from the various regular components 
of phase-space to the values of mean and variance. 
The study of this particular point is interesting because of the two 
following reasons. 
On the one hand, the method proposed in Refs.\ \cite{Rob93,Pro94} to compute 
the variance is not easy to use in practice since it requires the prior 
identification in phase-space of the adequate neighboring quantized tori. 
On the other hand, a given stable periodic orbit is in a way able all alone 
to account for the quantitative contribution of its associated regular 
component to the semiclassical density of states \cite{Vor75,Mil75,Gut90}. 
Considering these two facts, one may wonder whether stable periodic orbits 
(whose practical identification is usually easier than the one of specific 
quantized tori) cannot help to compute in a convenient way the quantitative 
contributions of the regular regions to the values of mean and variance. 
The paper shows that this is indeed the case by providing closed formulae for 
mean and variance in which stable periodic orbits are involved. 
Such formulae generalize to the case of mixed quantum systems those studied 
in Refs.\ \cite{Wil87,Eck92,Meh95,Boo95}. This work extends the one of 
Ref.\ \cite{Boo96} where details have been omitted because of lack of space.  
The illustrative numerical computations are again done in the framework of 
the Hydrogen atom in a strong magnetic field.

The paper is organized as follows. In Section II, some details relating to the
scaling property which characterizes the Hydrogen atom in a strong magnetic 
field are given and the choice of the perturbing operator used for the 
numerical computation of the transition matrix elements is justified. 
Section III deals with the mean value of the distribution of diagonal 
transition matrix elements, a quantity which was not considered in Refs.\
\cite{Rob93,Pro94}. 
The quantitative contributions of the ergodic and regular regions making up 
the mixed phase-space are identified. 
The contributions of the various regular components are calculated with the 
help of the associated stable periodic orbits. 
The illustrative numerical computations are done for three different values 
of the scaled energy. 
Section IV is concerned with the variance characterizing the distributions 
of non-diagonal transition matrix elements. 
First, the orders of magnitude of the transition probabilities whose initial
and final states are referring or not to the same phase-space component are 
compared. 
The quantitative contributions to the variance of the ergodic and regular 
regions are then studied. 
The manner in which a given stable periodic orbit is used in the calculation 
of the contribution coming from the associated regular component is discussed.
The illustrative numerical computations are done for the same values of the 
scaled energy as in the previous section. 
A Conclusion summarizes the main results of the paper.

\section{Scaling and transition matrix elements} 
To begin with, it is not an easy task to study the quantitative effect of the 
underlying classical dynamics on the statistical properties of transition 
matrix elements. 
This is due to the fact that the structure of phase-space changes with energy 
for most systems. 
Consequently, any diagonal transition matrix element 
$\langle m|\hat A|m \rangle$ ($|m \rangle$ being an eigenvector of the system 
and $\hat A$ the perturbing operator) is usually going with a different energy
shell. 
The problem becomes even trickier when considering non-diagonal transition 
matrix elements $\langle m|\hat A|n \rangle$ since initial and final 
eigenstates are always going with different energy shells. 
The study should therefore be done using a set of transition matrix elements 
in which initial and final eigenenergies are restricted to an energy interval 
small enough for phase-space to keep its structure practically unchanged.
However, in order to get a significant number of transition matrix elements, 
one should choose an energy region which is high enough in the spectrum so 
that the density of states can make up for the smallness of the used energy 
interval. 
This would obviously require the diagonalization of very large matrices in 
practice. 
Such numerical difficulties can be avoided by studying scaling systems. 
These are systems possessing scaling properties which imply that phase-space 
has the same structure at all energies. 
The Hydrogen atom in a strong magnetic field \cite{Fri89} and the three-body
Coulomb system \cite{Win92} are physically interesting examples of scaling 
systems. 
The Hamiltonian of a scaling system depends on a scaling parameter. 
It is useful to follow the variation of the energy levels with the scaling 
parameter \cite{Fri89,Win92} since one can extract informations about the 
underlying classical dynamics from spectra taken at different values of the 
scaling parameter by application of the so-called method of scaled energy 
spectroscopy \cite{Hol88,Mai94}. 

In atomic units, the quantum Hamiltonian of the Hydrogen atom in a strong 
magnetic field reads \cite{Fri89}
\begin{equation}
   \hat H_\gamma(\hat{\bf p},\hat{\bf r})
 = {1\over 2}\hat{\bf p}^2 - {1\over \hat r} + {1\over 2}\gamma \hat L_z
   + {1\over 8}\gamma^2({\hat x}^2+{\hat y}^2)  \; .
\label{Quantum}
\end{equation}
Here ${\hat L}_z$ is the component of the angular momentum operator along the
direction of the magnetic field. 
This component is conserved and, consequently, the azimuthal quantum number 
$m$ is a good quantum number. 
Numerical computations are restricted to the subspace $m = 0$ in this study.
The Hamiltonian is invariant under reflection with respect to the plane which 
is perpendicular to the direction of the magnetic field and the $z$-parity 
$\pi_z$ is thereby a good quantum number too.
The parameter $\gamma$ is the magnetic field strength in atomic units,
$\gamma=B/(2.35\times 10^5\,{\rm T})$.
When expressed in terms of the scaled coordinates 
$\tilde {\bf r} = \gamma^{2/3} {\bf r}$ and scaled momenta 
$\tilde {\bf p} = \gamma^{-1/3} {\bf p}$, the classical Hamiltonian scales as
\begin{eqnarray}
     \tilde H_{\gamma=1}(\tilde {\bf p},\tilde {\bf r}) 
 &=& {1\over2}\tilde {\bf p}^2 - {1\over \tilde r} + {1\over 2}\tilde L_z
     + {1\over 8}(\tilde x^2 + \tilde y^2)  \nonumber \\
 &=& \gamma^{-2/3}{H_{\gamma}}({\bf p},{\bf r}) = \gamma^{-2/3}E = \tilde E\; ,
\label{Hamscal}
\end{eqnarray}
with $E$ the excitation energy. 
Therefore, the classical dynamics obtained from the scaled equations of 
motion does not depend on $E$ and $\gamma$ independently but on a single
parameter combining both physical quantities, the scaled energy 
$\tilde E=\gamma^{-2/3}E$.
This implies that the structure of phase-space is identical for any pair 
($E$,$\gamma$) leading to the same value of the scaled energy.

If ${\hat{\tilde z}}$ and ${\hat{\tilde p}_{z}}$ are the scaled coordinate 
and momentum operators along the direction of the magnetic field, one has 
$[{\hat{\tilde z}},{\hat{\tilde p}_{z}}] = i\gamma^{1/3}\hbar$ by virtue of 
the previously given definition of scaled variables. 
The dependence of the quantum dynamics on the magnetic field strength 
$\gamma$ can thereby be taken into account by means of an effective Planck's 
constant ${\hbar}_{\rm eff}=\gamma^{1/3}\hbar$ ($\hbar=1$ subsequently). 
One approaches the semiclassical limit at constant scaled energy 
${\hbar}_{\rm eff} \rightarrow 0$ by decreasing the value of $\gamma$. 
The quantization of Eq.\ (\ref{Hamscal}) in the subspace $m = 0$ leads to a 
generalized eigenvalue equation for the scaling parameter 
$w = \gamma^{-1/3} = \hbar_{\rm eff}^{-1}$.
The introduction of the scaled semi-parabolic coordinates 
$\tilde \mu = \sqrt{{\tilde r}+{\tilde z}}$ and 
$\tilde \nu = \sqrt{{\tilde r}-{\tilde z}}$ allows to write this generalized 
eigenvalue equation as
\begin{eqnarray}
 && \left[2\tilde E ({\tilde \mu}^2+{\tilde \nu}^2)
   - {1 \over 4}{\tilde \mu}^2{\tilde \nu}^2({\tilde \mu}^2+{\tilde \nu}^2)
   + 4 \right] \, \Psi({\tilde \mu},{\tilde \nu}) \nonumber \\
 &=& w^{-2} \left({{\hat {\tilde p}}_\mu}^2 + {{\hat {\tilde p}}_\nu}^2\right) \,
   \Psi({\tilde \mu},{\tilde \nu})  \; ,
\label{H_eps}
\end{eqnarray}
with the radial operators ${{\hat {\tilde p}}_\mu}^2$ and 
${{\hat {\tilde p}}_\nu}^2$ defined as
\[
   {{\hat {\tilde p}}_\mu}^2
 = -{1 \over \tilde \mu}{\partial \over {\partial \tilde \mu}}
   \left(\tilde \mu{\partial \over {\partial \tilde \mu}}\right) \; , \quad
   {{\hat {\tilde p}}_\nu}^2
 = -{1 \over \tilde \nu}{\partial \over {\partial \tilde \nu}}
   \left(\tilde \nu{\partial \over {\partial \tilde \nu}}\right) \; .
\]
Eq.\ (\ref{H_eps}) can be written in matrix form by using the complete set of
basis functions which is composed of the tensorial products of the 
eigenvectors of two uncoupled two-dimensional harmonic oscillators with 
frequency $\sqrt{-2{\tilde E}}$ \cite{Fri89,Mai94b}.
The generalized eigenvalue equation is solved with the help of the Lanczos 
spectral transformation method, which is adapted to the diagonalization of 
sparse symmetric matrices \cite{Eri80}. 
One obtains in this way the spectrum of eigenvectors $|{{\Psi}_m} \rangle$ 
and corresponding eigenvalues $w_m = (\gamma^{-1/3})_m$. 
The structure of phase-space is the same at every eigenenergy $w_m$ since the 
generalized eigenvalue equation has been solved at constant scaled energy 
$\tilde E$. 
Consequently, the matrix elements describing the various transitions caused 
in the spectrum by a perturbing operator are all going with the same 
underlying classical dynamics.  
It has to be noted that the eigenvectors $|{{\Psi}_m} \rangle$ are not 
orthogonal.
However, the modified eigenvectors 
$|m \rangle = \sqrt{{{\hat {\tilde p}}_\mu}^2
 + {{\hat {\tilde p}}_\nu}^2}|{{\Psi}_m}\rangle$ 
(with the same corresponding eigenvalues $w_m$) are orthogonal, i.e.
\begin{equation}
   \langle m|n \rangle
 = \langle{\Psi}_m|{{\hat {\tilde p}}_\mu}^2
    + {{\hat {\tilde p}}_\nu}^2|{\Psi}_n\rangle = \delta_{mn}  \; .
\end{equation}
The perturbing operator which has been chosen for this study is
\begin{equation}
 \hat A = {1\over 2r\hat{\bf p}\,^2}
        = {w^2\over {{\hat {\tilde p}}_\mu}^2 + {{\hat {\tilde p}}_\nu}^2} \; .
\end{equation}
The second expression in this equation is the effective expression of the 
operator $\hat A$ when acting onto eigenvectors belonging to the subspace 
$m = 0$. 
The same operator has been used in previous papers \cite{Boo95,Boo96} already.
The reason is that this particular operator is convenient to handle in 
practice since the general expression of the associated transition matrix 
elements in the orthonormal basis $\lbrace |m \rangle \rbrace$ of modified
eigenvectors is 
\begin{equation}
   \langle m|\hat A|n \rangle
 = {w_m}{w_n}\langle {\Psi}_m|{\Psi}_n\rangle  \; .
\end{equation}
The computation of the transition matrix elements at constant scaled energy 
amounts therefore merely to the computation of the various overlaps of the 
eigenvectors $\lbrace |{\Psi}_m \rangle \rbrace$. 
Moreover, this computation can be restricted to a subspace which is labeled 
by a given value of $\pi_z$ because the operator $\hat A$ connects only 
eigenstates of the same $z$-parity. 
Numerical computations are done here in the subspace $m^{\pi} = 0^+$.
This special perturbing operator offers another computational advantage. 
Indeed, it turns out that the average at constant scaled energy of the Weyl 
transform $\tilde A$ of the operator $\hat A$ along a given trajectory in 
phase-space has an analytic expression, given in Eq.\ (\ref{averwe}) below. 
This average, which is important for the physical interpretation of the 
numerical results presented in this paper, can therefore be computed 
accurately.
The Weyl transform of the perturbing operator is expressed most simply with 
the help of the scaled semi-parabolic momenta 
${{\tilde p}_\mu}$ and ${{\tilde p}_\nu}$, the conjugate variables of the 
scaled semi-parabolic coordinates $\tilde \mu$ and $\tilde \nu$. 
These momenta are defined as the derivatives of the coordinates $\tilde \mu$ 
and $\tilde \nu$ with respect to the so-called rescaled time variable 
$\tau$ \cite{Fri89}.
The expression of $\tilde A$ is
\begin{equation}
   {\tilde A ({\tilde p}_\mu,{\tilde p}_\nu)}        
      = {1\over {{{\tilde p}_\mu}^2+{{\tilde p}_\nu}^2}}  \; .
\end{equation}
The scaling parameter $w = \gamma^{-1/3}$ and the scaled action $\tilde s
(\tilde E) = \int\,(\tilde p_\mu d\tilde \mu + \tilde p_\nu d\tilde \nu)$ 
are conjugate variables at constant scaled energy. 
Since the variable $w$ is an energy variable, the variable $\tilde s$
can be considered as a time variable which measures the length of trajectories
in the four-dimensional phase-space spanned by the scaled semi-parabolic 
variables $(\tilde\mu, \tilde\nu, \tilde p_\mu, \tilde p_\nu)$ \cite{Wif87}.
The infinitesimal scaled action $d\tilde s$ and the infinitesimal rescaled 
time $d\tau$ are linked together by the relation
\begin{equation} 
   d{\tilde s} = {{\tilde p}_\mu}d{\tilde \mu} + {{\tilde p}_\nu}d{\tilde \nu}
               = ({{\tilde p}_\mu}^2 + {{\tilde p}_\nu}^2)d{\tau}  \; .
\end{equation}
This relation implies that the Weyl transform $\tilde A$ can be expressed at
constant scaled energy as the derivative of the variable $\tau$ with respect 
to the variable $\tilde s$, i.e. 
\begin{equation}
   \tilde A(\tilde p_\mu,\tilde p_\nu)
 = {d\tau\over d\tilde s}
 = \tilde A(\tilde s) \; .
\end{equation}
The average at scaled energy ${\tilde E}$ of the Weyl transform $\tilde A$ 
along a trajectory with scaled action $S(\tilde E)$ corresponding to a total 
rescaled time $\tau$ is then simply
\begin{equation}
  {1\over S(\tilde E)}\int_0^{S(\tilde E)}d{\tilde s}{\tilde A(\tilde s)}
 = {1\over S(\tilde E)}\int_0^{\tau}d{\tau}
 = {{\tau}\over {S(\tilde E)}}  \; .
\label{averwe}
\end{equation}

\section{Diagonal transition matrix elements}
\begin{figure}[b]
\vspace{11.0cm}
\includegraphics{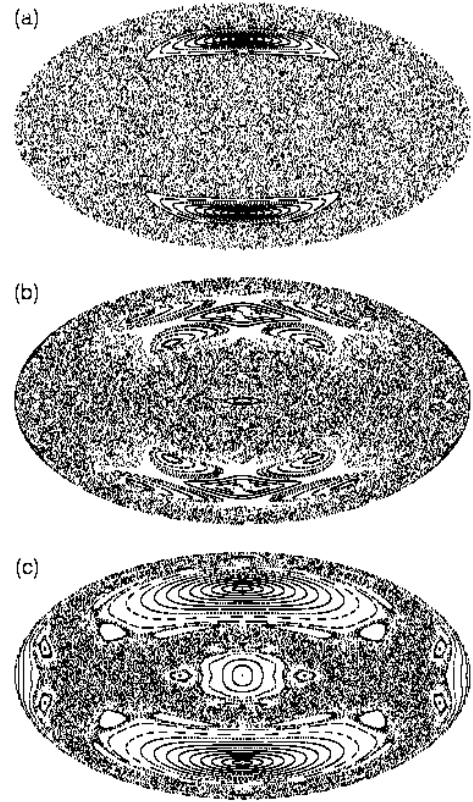}
\caption{\label{fig1} 
Poincar\'e surfaces of section of the Hydrogen atom in a strong magnetic 
field in the scaled semi-parabolic representation 
$(\tilde\mu, \tilde p_\mu; \tilde\nu=0)$ for three different values of the 
scaled energy $\tilde E$.
(a) $\tilde E=-0.2$;
(b) $\tilde E=-0.316$; (c) $\tilde E=-0.4$.}
\end{figure}

This section studies the quantitative contributions coming from the ergodic 
and regular components of the mixed phase-space to the mean value 
characterizing the distribution of diagonal transition matrix elements 
associated to the perturbing operator $\hat A$.
The study is done for three successive values of the scaling energy 
$\tilde E$, which are $\tilde E = -0.2$, $\tilde E = -0.316$ and 
$\tilde E= -0.4$. 
Fig.\ \ref{fig1} displays the Poincar\'e surfaces of section for these 
different cases in the scaled semiparabolic representation 
($\tilde\mu, \tilde p_\mu; \tilde\nu=0$).
In the case $\tilde E = -0.2$ (Fig.\ \ref{fig1}a), the surface of section 
has two islands of stable motion which are embedded in a simply connected 
ergodic sea. 
Both islands belong to the same regular component since they are conjugate 
with 
\newpage
\phantom{}
\begin{figure}[b]
\vspace{10.0cm}
\includegraphics{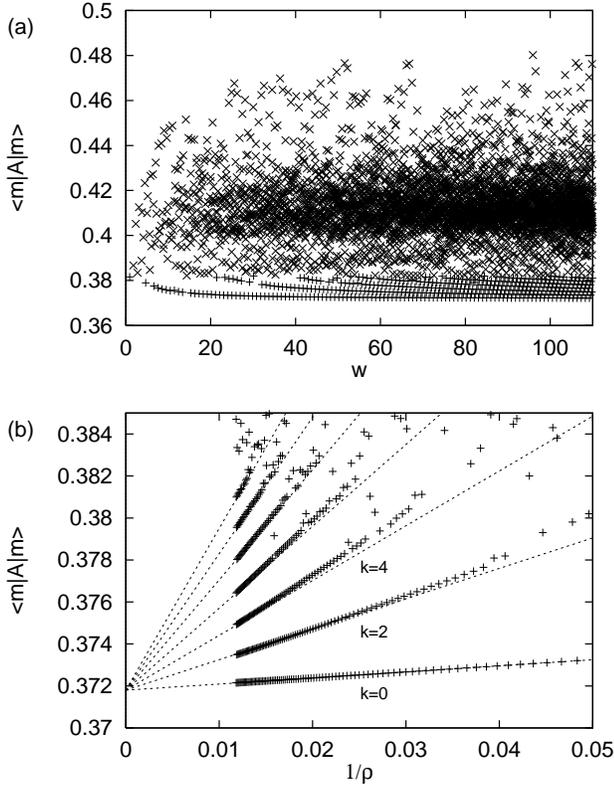}
\caption{\label{fig2} 
Distribution of values of the diagonal transition matrix elements 
$\langle m|\hat A|m\rangle$ 
corresponding to the operator $\hat A=1/2r\hat{\bf p}^2$ in the case of the 
scaled energy $\tilde E=-0.2$.
(a) Distribution of values of the matrix elements associated to the chaotic 
($\times$) and regular ({\tt +}) states as a function of the scaling 
parameter $w$;
(b) Distribution of values of the matrix elements associated to the regular 
states as a function of the inverse of the mean total density of states
$\rho_{0,t}(w)$.}
\end{figure}
\noindent
respect to the line $\tilde p_\mu=0$.
This regular component is associated to a stable periodic orbit lying in the 
plane perpendicular to the direction of the magnetic field, with scaled action
$S_r = 6.49086$. 
Fig.\ \ref{fig2}a displays, for the same value of the scaling energy, the 
distribution of values of the 
diagonal transition matrix elements as a function of the
scaling parameter $w = \gamma^{-1/3} = {\hbar}_{\rm eff}^{-1}$. 
This figure shows clearly the presence of two different patterns. 
On the one hand, there appears a statistical distribution built up with 
diagonal transition matrix elements which are represented by "$\times$" 
symbols. 
A further analysis reveals that these matrix elements involve eigenstates 
whose Husimi distribution \cite{Mue94} is essentially localized within the
ergodic sea. 
Such eigenstates are called ``chaotic'' in the sequel. 
They are generically labelled as $|m_c \rangle$, with corresponding 
eigenvalues $w_m^{(c)}$. 
These eigenvalues belong to a subset of the spectrum which is characterized 
by a mean density of states ${\rho}_{0,c}(w) = 0.702w$. 
On the other hand, one distinguishes in the lower part of Fig.\ \ref{fig2}a 
several sequences of diagonal matrix elements which are represented by 
"{\tt +}" symbols.
The Husimi distributions of the corresponding eigenstates are well localized
within the two islands of stable motion in Fig.\ \ref{fig1}a.
Such eigenstates are called ``regular'' in the sequel. 
They are generically labelled as $|m_r \rangle$, with corresponding 
eigenvalues $w_m^{(r)}$. 
These eigenvalues belong to a subset of the spectrum which is complementary  
to the previous one and whose mean density of states is
${\rho}_{0,r}(w) = 0.066w$. 
The coexistence of these two different patterns of diagonal transition matrix 
elements illustrates well the fact that the spectrum of a mixed quantum system
can be split into independent subsets of eigenvectors, each of which going 
with a different component of phase-space \cite{Bro84}. 

The quantitative interpretation of Fig.\ \ref{fig2}a is done with the help of 
a scaled spectral function $D(w)$ which takes all diagonal matrix elements of 
the perturbing operator $\hat A$ into account. It is defined as 
\begin{equation}
\label{suru}
 D(w) = \sum_m{\langle m|{\hat A}|m \rangle}{\delta(w-w_m)} \; .
\end{equation}
According to Refs.\ \cite{Wil87,Eck92}, the semiclassical expression of 
$D(w)$ which is relevant to the fully ergodic case is
\begin{eqnarray}
     D(w)
 &=& {w^4\over(2\pi)^2}
 \int_{{\Omega}} d{{\tilde p}_\mu} d{{\tilde p}_\nu} d{\tilde \mu} d{\tilde \nu} \, 
 \left({{\tilde \mu}^2+{\tilde \nu}^2}\right) \nonumber \\
 &\times& {\tilde A({\tilde p}_\mu, {\tilde p}_\nu)}\, 
 {\delta({\tilde H({{\tilde p}_\mu}, {{\tilde p}_\nu}, {\tilde \mu}, 
 {\tilde \nu})}-2)}  \nonumber \\
 &+& {1\over \pi }\sum_{\{p\}} A_p\,\sum_{s=1}^\infty
 {{\sin\left(s\left({w S_p(\tilde E)}
 -{\alpha_p\pi/2}\right)\right)}\over {\left[\det (M_p^s - I)\right]^{1/2}}}  \; ,
\label{diagonal}
\end{eqnarray}
with
\begin{eqnarray}
 & & {\tilde H({{\tilde p}_\mu}, {{\tilde p}_\nu}, {\tilde \mu}, {\tilde \nu})}
     \nonumber \\
 &=& {1\over 2} \left({{\tilde p}_\mu}^2+{{\tilde p}_\nu}^2\right)
    -({\tilde \mu}^2+{\tilde \nu}^2) \tilde E \,
    +{1\over 8}{\tilde \mu}^2{\tilde \nu}^2({\tilde \mu}^2+{\tilde \nu}^2) \; .
\label{H_semi}
\end{eqnarray}
The expression of $D(w)$ is written here in the scaled semi-parabolic 
representation of phase-space.
The first term on the r.h.s.\ of Eq.\ (\ref{diagonal}), the so-called Weyl 
term, gives the classical contribution to the scaled spectral function. 
It factorizes into two parts as a consequence of the use of the scaled 
semiparabolic representation. 
One part depends solely on the scaling parameter $w$ whereas the other part,
which involves a phase-space integration over the whole energy shell 
$\Omega$ at 
constant scaled energy $\tilde E$, is independent of it. 
The second term gives the leading order corrections to the Weyl term in an
asymptotic expansion of $D(w)$ into powers of $\hbar$. These corrections are 
generated by the periodic orbits of the underlying classical dynamics, which 
are all unstable. The outer sum in the second term runs over the set 
$\{p\}$ of all primitive periodic orbits whereas the inner sum runs over all 
repetitions $s$ of every primitive periodic orbit. Each primitive periodic 
orbit $p$ is characterized by an amplitude $A_p$ whose expression is
\begin{equation}
\label{ampp}
 A_p = \oint d\tilde s \, \tilde A(\tilde s) \; . 
\end{equation}
In this equation, the integration is done along the considered periodic orbit 
and over one period 
$S_p(\tilde E) = \oint (\tilde p_\mu d\tilde\mu + \tilde p_\nu d\tilde\nu)$.
The scaled action $S_p(\tilde E)$ appears in the argument of the sine function
together with the scaling parameter $w$ since it scales as 
$S_p(\tilde E) = w^{-1}S_p(E)$, $S_p(E)$ being the usual action computed at 
the excitation energy $E = \tilde E/w^2$.
Every primitive periodic orbit is also characterized by its monodromy matrix 
$M_p$ and its Maslov index ${\alpha}_p$, which are both independent of energy
for scaling systems. The importance of the correction due to the $s$-th 
repetition of the primitive periodic orbit $p$ depends on the determinant of 
the difference between the $s$-th power of $M_p$ and the unit matrix $I$. 
As Fig.\ \ref{fig2}a shows, the scaled spectral function $D(w)$ can be written 
as the sum of two different contributions in the case of a mixed system, i.e.,
\begin{equation}
\label{csum}
 D(w) = D_{c}(w) + D_{r}(w)
\end{equation}
with
\begin{equation}
\label{subsum}
 D_{c(r)} = \sum_{m_{c(r)}} 
 \langle m_{c(r)}|\hat A|m_{c(r)}\rangle \, \delta(w - w_m^{(c(r))})  \; .
\end{equation}
The semiclassical expression of $D_{c}(w)$ (resp.\ $D_{r}(w)$), the scaled 
spectral function which is associated to the ergodic (resp.\ regular)
component of phase-space, is similar to the one given in Eq.\ (\ref{diagonal}).
The integration in the Weyl term is now done over the corresponding part of the
energy shell at constant scaled energy. The leading order corrections to the
Weyl term are due to the unstable $(D_{c}(w))$ or stable $(D_{r}(w))$ periodic
orbits of the underlying classical dynamics.  

The semiclassical expression of $D_{c}(w)$ predicts that the classical
contribution to the mean value $\langle \hat A\rangle_c$ which characterizes
the distribution of diagonal transition matrix elements corresponding to 
the chaotic
eigenstates is given by the microcanonical average of the operator $\hat A$ 
over the ergodic part ${\Omega}_c$ of the energy 
shell \cite{Wil87,Eck92}, i.e.,
\begin{equation}
\label{miav} 
 \langle \hat A\rangle_c = 
 {{\int_{{\Omega}_c} d{{\tilde p}_\mu} d{{\tilde p}_\nu} d{\tilde \mu} d{\tilde \nu} \, 
 \left({{\tilde \mu}^2+{\tilde \nu}^2}\right)
 {\tilde A({\tilde p}_\mu, {\tilde p}_\nu)}\,
 {\delta({\tilde H}-2)}} \over
 {\int_{{\Omega}_c} d{{\tilde p}_\mu} d{{\tilde p}_\nu} d{\tilde \mu} d{\tilde \nu} 
 \left({\tilde \mu}^2+{\tilde \nu}^2\right) 
 {\delta({\tilde H}-2)}}} \; ,
\end{equation}
with $\tilde H({{\tilde p}_\mu}, {{\tilde p}_\nu}, {\tilde \mu}, {\tilde \nu})$
defined in Eq.\ (\ref{H_semi}) \cite{Boo96:Erratum}.
This formula is not easy to use in practice for the computation of the 
classical contribution since it requires the numerical calculation of 
three-dimensional phase-space 
integrals. However, just as in Refs.\ \cite{Meh95,Boo95}, the ergodic theorem 
allows to replace the microcanonical average with a time (i.e.\ scaled action)
average of the Weyl transform of the perturbing operator along any trajectory 
exploring the ergodic part of the energy shell in a uniform way. 
With the help of Eq.\ (\ref{averwe}), the classical contribution to the mean
value $\langle \hat A\rangle_c$ can be finally expressed as  
\begin{equation}
\label{erav}
   \langle \hat A\rangle_c
 = \lim_{S\to\infty} {1\over S} \int_0^S d\tilde s \, \tilde A(\tilde s) 
 = \lim_{S\to\infty} {\tau(S)\over S} \; ,
\end{equation}
$\tau(S)$ being the total rescaled time corresponding to the scaled action $S$.
The ergodic theorem ensures that the limit on the r.h.s.\ of this equation
is well defined and unique. This formula is well suited for the numerical
calculation of the classical contribution to the mean value. In the case 
$\tilde E = -0.2$, it gives the value $\langle \hat A\rangle_c = 0.41$, 
which is in good agreement with the value 0.411 obtained from the statistical
distribution of Fig.\ \ref{fig2}a. Eq.\ (\ref{erav}) can therefore be used to 
compute with an excellent precision the mean value characterizing the 
distribution of diagonal matrix elements associated to the chaotic eigenstates.

Fig.\ \ref{fig2}a suggests that all sequences of diagonal matrix which 
correspond to the regular eigenstates are converging towards the same limit  
as the scaling parameter $w$ increases. This is indeed the case, as clearly 
shown by Fig.\ \ref{fig2}b. In this figure, the values of the diagonal  
matrix elements pertaining to the seven sequences of the
previous figure are represented as a function of the inverse of the mean total
density of states $\rho_{0,t}(w) = \rho_{0,c}(w) + \rho_{0,r}(w)$. Such a  
representation is useful since it allows to extrapolate the values of the 
individual diagonal matrix elements in the semiclassical regime. This comes 
from the fact that the mean total density of states is proportional to $w$ 
in the case of the Hydrogen atom in a strong magnetic field
($\rho_{0,t} = 0.768w$ in the case $\tilde E = -0.2$), and so its inverse is 
proportional to the effective Planck's constant 
$\hbar_{\rm eff} = \gamma^{1/3}$ of this system. The
semiclassical limit at constant scaled energy $\hbar_{\rm eff}\to 0$ is 
therefore reached as $1/\rho_{0,t}(w)\to 0$. One sees that the values of    
the diagonal matrix elements in the sequences are approximated all the better
by a common limiting value $\langle \hat A\rangle_r$ as one comes closer to  
the semiclassical regime. This behavior is in sharp contrast with the one of 
the diagonal matrix elements corresponding to the chaotic eigenstates, which 
are distributed statistically around the mean value $\langle \hat A\rangle_c$. 
The limiting value $\langle  \hat A\rangle_r$ can be most easily computed by
using an expression of the scaled spectral function $D_{r}(w)$ which is
obtained by resumming formally the leading order corrections to the Weyl term 
generated by the stable periodic orbit and all its repetitions in the 
following way \cite{Mil75,Gut71,Gut90}. The expression of these corrections 
for a particular stable periodic orbit $r$ is \cite{Mil75,Gut71,Gut90} 
\begin{equation}
\label{misu}
 -{i\over \pi}\,A_r \sum_{s=1}^\infty 
  {\exp(is(w S_r(\tilde E) - \alpha_r\pi/2)) \over 2\sin(s\pi\gamma_r)} \; .
\end{equation}
Here ${\gamma}_r$ is the winding number of the stable orbit, which is  
independent of energy for scaling systems. After expansion of the sine function
in the denominator and the subsequent use of the Poisson summation 
formula, this expression can be rewritten as 
\begin{equation}
\label{pois}
 A_r \sum_{n,k}{\delta}\left(w S_r(\tilde E) 
 - 2\pi\left(n + {\alpha_r\over 4}\right) 
 - 2\pi\left(k + {1\over 2}\right)\gamma_r\right) \; .
\end{equation}
This is a formal expression of the scaled spectral function $D_r(w)$ in the
semiclassical regime. It predicts that the semiclassical spectrum of 
the regular states is similar to the spectrum of a two-dimensional harmonic 
oscillator \cite{Mil75,Gut71,Gut90}. One direction of harmonic motion is 
along the stable periodic orbit, whereas the other direction of harmonic 
motion is transverse to it. This motion takes place on neighboring tori 
surrounding 
the orbit. Each such torus is identified by two quantized scaled actions, the
action ${\tilde I}_{n} = 2\pi(n + \alpha_r/4)$ which is associated to the 
longitudinal motion and the action ${\tilde I}_{k} = 2\pi(k + 1/2)\gamma_r$ 
which is associated to the transverse motion \cite{Per77}. Every regular state 
is therefore labelled by two quantum numbers $n$ and $k$. The range of values 
of both quantum numbers is bounded in practice as a consequence of the finite 
size of the corresponding region of stable motion \cite{Boh93}. It has been 
checked that all  
regular eigenstates building up a given sequence of diagonal matrix elements 
in Figs.\ \ref{fig2}a,b are labelled by the same quantum number $k$. This 
quantum number is even as a result of the parity symmetry with respect to the 
plane perpendicular to the direction of the magnetic field. The lowest 
sequence corresponds to $k=0$, the next one to $k=2$ and so on until the last
identified sequence which corresponds to $k=12$. Fig.\ \ref{fig2}b shows that
there are a few diagonal matrix elements which do not belong to any sequence.
It has been checked that all of them correspond to regular eigenstates which
are involved in quasi-crossings and so cannot be labelled by a fixed value of 
$k$. The quantum number $n$ differentiates the regular eigenstates which are 
associated to a given sequence. The previous formal expression of the scaled 
spectral function $D_r(w)$ predicts also that the limiting value  
$\langle \hat A\rangle_r$ of the sequences is proportional to the amplitude 
$A_r$ of the stable periodic orbit, whose expression is given by 
Eq.\ (\ref{ampp}). The amplitude $A_r$ has to be divided by the scaled action 
$S_r$ of the orbit for the purpose of normalization. With the help of 
Eq.\ (\ref{averwe}), this limiting value can be expressed as
\begin{equation}
\label{avreg}
 \langle {\hat A} \rangle_r
 = {1\over {S_r}} \oint d{\tilde s}\, \tilde A(\tilde s) 
 = {{\tau}_r\over {S_r}} \; ,
\end{equation}
${\tau}_r$ being the total rescaled time corresponding to the scaled action
$S_r$. It has to be remarked that the classical contribution 
$\langle \hat A\rangle_c$ to the mean characterizing the distribution of
diagonal matrix elements associated to the chaotic 
eigenstates, Eq.\ (\ref{erav}), and the classical limiting value 
$\langle \hat A\rangle_r$ of the sequences of diagonal matrix elements 
corresponding to the regular eigenstates, Eq.\ (\ref{avreg}), have similar
expressions. Each expression makes use of the classical trajectory which is
related to the component of phase-space going with the studied subset 
of diagonal matrix elements. In the case $\tilde E = -0.2$, the scaled action
corresponds to a total rescaled time ${\tau}_r = 2.4135$. This gives  
the value $\langle \hat A \rangle_r = 0.372$ which, as seen in 
Fig.\ \ref{fig2}b, agrees very well with the value obtained as the 
common intersection of the dashed lines interpolating the values of the 
diagonal matrix elements as one goes towards the semiclassical limit.  

Since the total distribution of diagonal matrix elements is composed of two 
different subsets corresponding to the two different components of 
phase-space, its mean value $\langle \hat A\rangle_t$ is expected to be the  
sum of two properly weighted contributions. The contribution which comes from 
the mean value $\langle \hat A\rangle_c$ (resp.\ limiting value 
$\langle \hat A\rangle_r$) is weighted by the ratio of the mean density of 
states of the subset of chaotic (resp.\ regular) states to the mean total 
density of states. Each weighting factor is roughly equal to the relative
volume of the corresponding component of phase-space. The expected expression 
of $\langle \hat A\rangle_t$ is therefore the following
\begin{equation}
\label{centro}
 {\langle{\hat A}\rangle_t}  
 =\left({{\rho_{0,c}}(w)\over {\rho_{0,t}}(w)}\right){\langle{\hat A}\rangle_c}
 +\left({{\rho_{0,r}}(w)\over {\rho_{0,t}}(w)}\right){\langle{\hat A}\rangle_r} \; .
\end{equation}
In the case $\tilde E = -0.2$, this formula gives the value 
$\langle{\hat A}\rangle_t = 0.407$, which is very close from the value
$\langle{\hat A}\rangle_t = 0.409$ obtained from the whole distribution of
Fig.\ \ref{fig2}a. Consequently, Eq.\ (\ref{centro}) allows to calculate with 
an excellent precision the mean value characterizing the total distribution 
of diagonal matrix elements. It generalizes to the case of a scaled system 
with a mixed phase-space the expression of the mean value given in 
Refs.\ \cite{Wil87,Eck92} for a scaled system with an ergodic phase-space. 

\begin{figure}[b]
\vspace{10.0cm}
\includegraphics{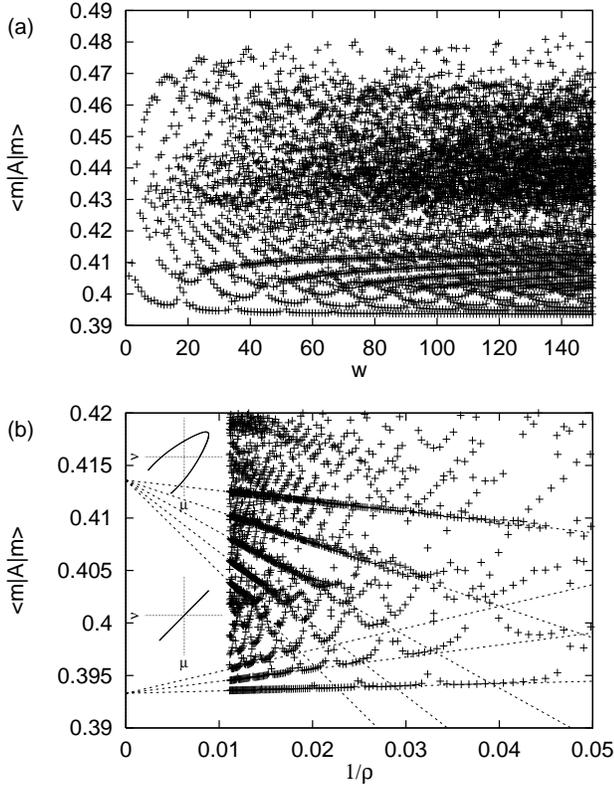}
\caption{\label{fig3} 
Same as Fig. 2 but in the case of the scaled energy $\tilde E=-0.316$.
The shapes of the two stable periodic orbits which are used in the
analysis of this case are shown in the insets.}
\end{figure}

It is expected that there are as many limiting values $\langle \hat A\rangle_r$
as there are regular components in phase-space. This important point can be 
checked by studying the distribution of diagonal matrix elements
for values of the scaled energy for which several stable periodic orbits are
present in the underlying classical dynamics. This is done in the cases 
$\tilde E = -0.316$ and $\tilde E = -0.4$. The Poincar\'e surface of section 
corresponding to the first case is displayed in Fig.\ \ref{fig1}b. There are  
five islands of stable motion on both sides of the line $\tilde p_\mu=0$,
a central island surrounded by a chain of four islands. The two conjugate 
central islands belong to the same regular component which, as in the case
$\tilde E = -0.2$, is associated to a stable periodic orbit lying in the plane
perpendicular to the direction of the magnetic field. This orbit has a scaled
action $S_{r1} = 7.903534$, corresponding to a total rescaled time
${\tau}_{r1} = 3.108460$. The shape of this orbit is depicted in the 
semi-parabolic coordinate representation $(\tilde\mu, \tilde\nu)$ in the lower
inset of Fig.\ \ref{fig3}b. The two conjugate chains of four islands of stable
motion surrounding the central islands belong to another regular component.
The associated stable periodic orbit has a scaled action $S_{r2} = 12.158133$,
corresponding to a total rescaled time ${\tau}_{r2} = 5.028604$. The shape of
this orbit is depicted in the scaled semi-parabolic coordinate representation
$(\tilde\mu, \tilde\nu)$ in the upper inset of Fig.\ \ref{fig3}b. 
Fig.\ \ref{fig1}c shows the Poincar\'e surface of section corresponding to the
case $\tilde E = -0.4$. As in the two previous cases, the two conjugate 
islands of stable motion on both sides of the line $\tilde p_\mu=0$
belong to a regular component with a stable periodic orbit lying in the plane
perpendicular to the direction of the magnetic field. This orbit has a scaled
action $S_{r1} = 7.02481$ corresponding to a total rescaled time 
${\tau}_{r1} = 2.866$. The stable island in the middle of the surface of 
section pertains to 
\newpage
\phantom{}
\noindent
another regular component, with a stable periodic orbit
which is parallel to the direction of the magnetic field. This orbit has a
scaled action $S_{r2} = 5.791216$, corresponding to a total rescaled time
${\tau}_{r2} = 2.896$.

\begin{figure}[b]
\vspace{10.0cm}
\includegraphics{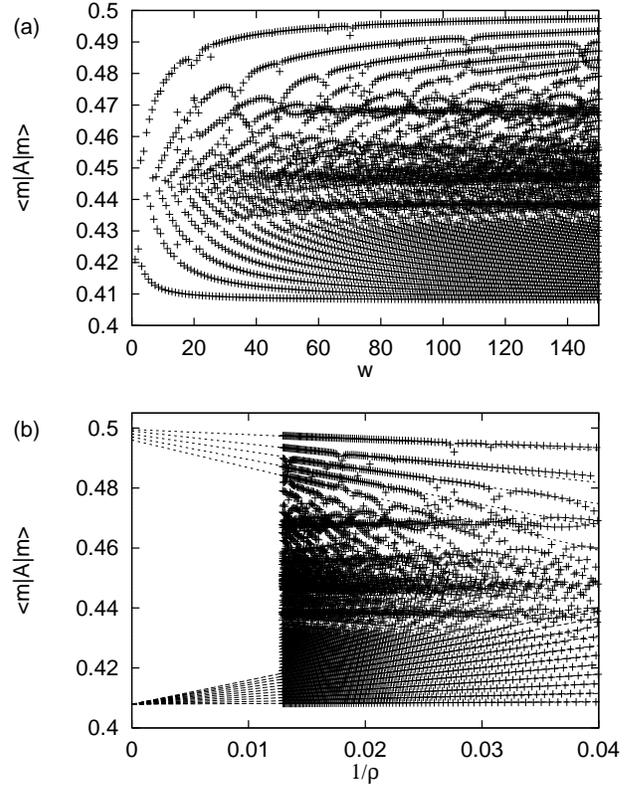}
\caption{\label{fig4} 
Same as Fig. 2 but in the case of the scaled energy $\tilde E=-0.4$.}
\end{figure}

Fig.\ \ref{fig3}a (resp.\ Fig.\ \ref{fig4}a) displays the distribution  
of values of the
diagonal matrix elements as a function of the scaling parameter $w$ in the 
case $\tilde E = -0.316$ (resp.\ $\tilde E = -0.4$). Just as 
in the case $\tilde E = -0.2$, the whole distribution results from the 
juxtaposition of a statistical distribution connected to the ergodic component 
of phase-space and a set of sequences connected to the regular components of
phase-space. These sequences fall into two different groups, as the 
fingerprint of the existence of two different stable periodic orbits. This
is particularly clear in Fig.\ \ref{fig4}a, where each group of 
sequences is on a different side of the statistical distribution. In the case
$\tilde E = -0.4$, the long streaks of diagonal matrix elements appearing in 
the statistical distribution are presumably the mark of one or several other
regular components, which are associated to the small islands of stable motion
surrounding the three main islands in Fig.\ \ref{fig1}c. As previously 
explained, the identification of the group of sequences linked to a
particular regular component is most easily done by representing the values 
of the diagonal matrix elements in all sequences as a function of the inverse
of the mean total density of states $\rho_{0,t}(w)$. This 
\newpage
\phantom{}
\noindent
is done in 
Fig.\ \ref{fig3}b (resp.\ Fig.\ \ref{fig4}b) for the case $\tilde E = -0.316$
(resp.\ $\tilde E = -0.4$). In Fig.\ \ref{fig3}b, one sees that the three lower
and the five upper dashed lines interpolating the values of the diagonal
matrix elements as one goes towards the semiclassical limit have a common 
intersection. The value of the common intersection of the three lower 
sequences is very precisely equal to the limiting value 
$\langle {\hat A}\rangle_{r1} = {{\tau}_{r1}/S_{r1}} = 0.3933$ of the 
stable periodic orbit lying in the plane perpendicular to the direction of the
magnetic field. The regular component which is related to these sequences is
therefore identified in a straightforward way. It is obviously the one to 
which the two conjugate central islands of stable motion of Fig.\ \ref{fig1}b
are pertaining. As in the case $\tilde E = -0.2$, the three lower sequences 
are labelled by even values of the quantum number $k$, with the lowest
sequence corresponding to $k=0$. The value of the common intersection of the 
five upper sequences is very precisely equal to the limiting value 
$\langle {\hat A}\rangle_{r2} = {{\tau}_{r2}/S_{r2}} = 0.4136$ of the other 
stable periodic orbit. Consequently, the regular component which is associated
to these sequences is the one whose fingerprint in Fig.\ \ref{fig1}b
 is constituted by the two conjugate chains of four islands of stable motion. 
Since this other orbit has no symmetry with respect to the plane perpendicular 
to the direction of the magnetic field, the five upper sequences are labelled 
by both even and odd values of the quantum number $k$, with the uppermost 
sequence corresponding to $k=0$. Similar results are obtained in the case 
$\tilde E = -0.4$. Indeed, one sees in Fig.\ \ref{fig4}b that the groups of 
dashed interpolating lines corresponding to the sequences above and below the 
statistical distribution have each a common intersection. The value of the
common intersection of the sequences below the statistical distribution is
very precisely equal to the limiting value 
$\langle {\hat A}\rangle_{r1} = {{\tau}_{r1}/S_{r1}} = 0.408$ of the stable
periodic orbit lying in the plane perpendicular to the direction of the
magnetic field. The regular component connected to these sequences is, as 
in the two previous cases, the one to whom the two conjugate islands of stable
motion in the Poincar\'e surface of section are belonging. This means that the
limiting value of a group of sequences readjusts itself to the change in 
length of the associated orbit as one moves from a particular example of the
mixed phase-space to another. The sequences below the statistical 
distribution are labelled by even values of $k$, with the lowest sequence
corresponding to $k=0$. As expected, the value of the common intersection of
the sequences above the statistical distribution is (nearly) equal to the 
limiting value $\langle {\hat A}\rangle_{r2} = {{\tau}_{r2}/S_{r2}} = 0.500$ 
of the stable periodic orbit which is parallel to the direction of the 
magnetic field. By way of consequence, the regular component which is linked 
to these sequences is the one whose fingerprint in Fig.\ \ref{fig1}c is the 
central island of stable motion. These sequences are labelled by both even and
odd values of $k$, with the uppermost sequence corresponding to $k=0$.  
Finally, the formula in Eq.\ (\ref{centro}) giving the mean value which 
characterizes the complete distribution of diagonal matrix elements is  
generalized to the case of several regular components simply by adding as many 
weighted classical values $\langle {\hat A}\rangle_r$ as there are stable
periodic orbits in the underlying classical dynamics.

\section{Non-diagonal transition matrix elements}
This section studies the quantitative contributions of the ergodic and regular
components of the mixed phase-space to the variance characterizing the 
distributions of non-diagonal transition matrix elements associated to the 
perturbing operator $\hat A$. This is done with the help of scaled spectral 
functions which take the different types of non-diagonal transition matrix 
elements into account, those coupling chaotic or regular 
eigenstates together as well as those coupling a chaotic and a regular 
eigenstate together. Since they deal with non-diagonal matrix elements, these 
scaled spectral functions depend necessarily on two energy scales, the scaled 
energy $w$ and the scaled energy difference $\Delta w$. The scaled total 
spectral function $C(w,\Delta w)$ is defined to be the sum of these scaled 
spectral functions, i.e., 
\begin{eqnarray}
     C(w,\Delta w)
  &=& C_{cc}(w,\Delta w)+C_{rr}(w,\Delta w) \nonumber \\
  &+& C_{cr}(w,\Delta w)+C_{rc}(w,\Delta w)
\label{Specto}
\end{eqnarray}
with
\begin{eqnarray}
 && C_{\alpha\beta}(w,\Delta w) \nonumber \\
   &=& \sum_{m_\alpha,n_\beta}
   {|\langle m_\alpha|\hat A - {\langle {\hat A}\rangle_{\alpha}}
   {\delta_{{\alpha}{\beta}}} |n_\beta\rangle|^2} 
   \delta_\eta\left(w-{{w_m^{(\alpha)}}+{w_n^{(\beta)}}\over 2}\right)
   \nonumber \\
   &\times&  \delta_\epsilon
   \left(\Delta w-\left(w_n^{(\beta)}-w_m^{(\alpha)}\right)\right) 
\label{Coirreg}
\end{eqnarray}
and $\alpha=c,r$; $\beta=c,r$. The (Lorentzian) smoothings of the 
$\delta$-functions, of widths $\eta$ and $\epsilon$, are introduced so that 
one can compute these spectral functions in practice in spite of the 
discreteness of the spectrum. The values of the widths which have been chosen
for the numerical computations are $\eta = 5.0$ and $\epsilon = 0.02$, in units
of the scaling parameter $w$. The classical values $\langle \hat A\rangle_c$,
$\langle \hat A\rangle_r$ are subtracted from the appropriate spectral 
functions in order to eliminate the quantitative contribution of the 
diagonal matrix elements in the semiclassical regime. The local
variance $\sigma_c^2(w,\Delta w)$ which is associated to the statistical
distributions of non-diagonal matrix elements coupling chaotic 
eigenstates together is related to the spectral function $C_{cc}(w,\Delta w)$ 
and to the mean density of states $\rho_{0,c}(w)$ by the formula \cite{Wil87}
\begin{equation}
           \sigma_c^2(w,\Delta w) 
           = {C_{cc}(w,\Delta w)\over \left(\rho_{0,c}(w)\right)^2}  \; .
\label{Varcha}
\end{equation}
By analogy with this formula, the local variance 
$\sigma_t^2(w,\Delta w)$ which
is associated to the distributions of all non-diagonal matrix elements is 
defined through the expression 
\begin{equation}
           \sigma_t^2(w,\Delta w) 
           = {C(w,\Delta w)\over \left(\rho_{0,t}(w)\right)^2}  \; .
\label{Var}
\end{equation}
The quantitative contribution of each spectral function in Eq.\ (\ref{Specto})
to the variance $\sigma_t^2(w,\Delta w)$ is studied in the sequel.

\begin{figure}[b]
\vspace{9.8cm}
\includegraphics{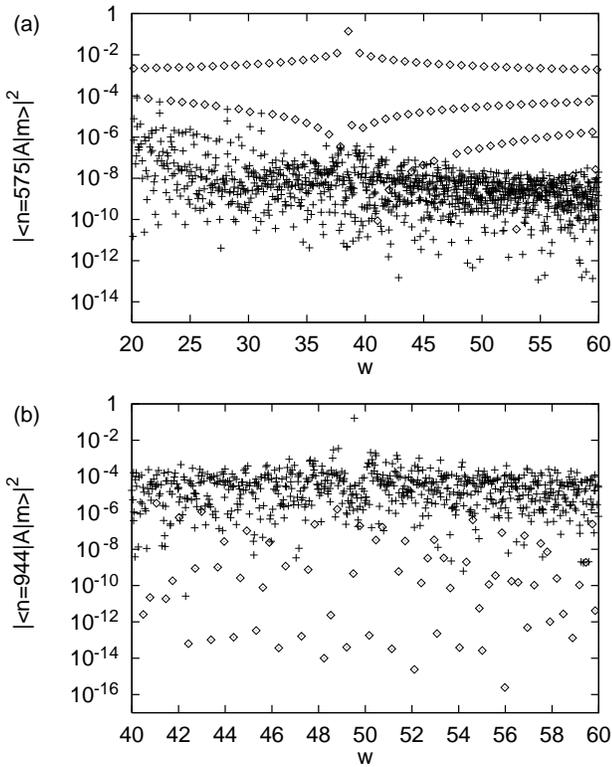}
\caption{\label{fig5} 
Distribution of values of the transition probabilities 
$|\langle n|\hat A|m\rangle|^2$ as a function of the scaling parameter $w$
in the case of the scaled energy $\tilde E=-0.2$.
(a) Regular eigenstate $|n\rangle$; 
(b) Chaotic eigenstate $|n\rangle$.
The diamonds and crosses mark the transitions to regular and chaotic states
$|m\rangle$, respectively.} 
\end{figure}

The spectral function $C_{cc}(w,\Delta w)$ (resp.\ $C_{rr}(w,\Delta w)$) is 
connected to the regular (resp.\ ergodic) component of phase-space in a manner 
which is detailed below. On the contrary, the spectral functions  
$C_{cr}(w,\Delta w)$ and $C_{rc}(w,\Delta w)$ cannot be associated to a 
particular component of phase-space since they take both chaotic and regular
eigenstates into account. According to Refs.\ \cite{Rob93,Pro94}, the 
quantitative contributions of these two last spectral functions to
$C(w,\Delta w)$ should therefore be significantly smaller than those of the 
two first spectral functions. Fig.\ \ref{fig5} shows that this is indeed the
case in practice. This figure displays the distribution of values of the 
transition probabilities $|\langle n|\hat A|m\rangle|^2$ from a given initial
state $|n\rangle$ to a subset {$|m\rangle$} of final eigenstates 
corresponding to a finite range of the energy spectrum at scaled energy
$\tilde E = -0.2$. In Fig.\ \ref{fig5}a, the chosen eigenstate $|n\rangle$
(the $575^{\rm th}$ state above the groundstate) is a regular eigenstate
with $k=0$ and eigenvalue $w_n^{(r)} \simeq 38.5$. On the contrary, the  
eigenstate $|n\rangle$ in Fig.\ \ref{fig5}b
(the $944^{\rm th}$ state above the groundstate) is a chaotic eigenstate.
In both figures, the diamonds (resp.\ crosses) are recording the values of the 
\newpage
\phantom{}
\noindent
transition probabilities when the final eigenstate $|m\rangle$ is a 
regular (resp.\ chaotic) one. Each of the three visible sequences of diamonds 
in Fig.\ \ref{fig5}a is associated to regular eigenstates which are labelled 
by the same (even) value of $k$, with the uppermost sequence corresponding to 
$k=0$. This figure shows clearly   
that the probabilities of most transitions coupling regular eigenstates 
together are several orders of magnitude larger than those coupling regular 
and chaotic eigenstates together. This implies in return that the values of 
the spectral function $C_{rr}(w,\Delta w)$ are much larger that those of the 
spectral functions $C_{rc}(w,\Delta w)$ and $C_{cr}(w,\Delta w)$. Two other 
observations in Fig.\ \ref{fig5}a are of interest. The first is that the 
square
of the diagonal matrix element $\langle n = 575|\hat A|n = 575\rangle$ is at
least one order of magnitude larger than the probabilities of transitions
coupling different regular eigenstates together. The second observation is 
that the probabilities of the transitions involving regular eigenstates 
with the same value of $k$ are more than one order of magnitude larger
than those involving regular eigenstates with different values of $k$. 
Moreover, the larger the difference in $k$ between two regular eigenstates, 
the smaller the corresponding transition probability \cite{Per77}. It has 
also been checked that the values of the transition probabilities belonging to 
the sequence of diamonds with $k=0$ are decaying exponentially as one goes
away from the diagonal matrix element. This type of decay is a universal 
feature of one-dimensional systems \cite{Landau}. The system behaves therefore
as an effective one-dimensional system at the level of the transition 
amplitudes which are connected to the first quantized torus surrounding the 
stable orbit. Fig.\ \ref{fig5}b shows clearly that the probabilities of most
transitions involving chaotic eigenstates are also several orders of 
magnitude larger than those involving chaotic and regular eigenstates. As a
consequence, the values of the spectral function $C_{cc}(w,\Delta w)$ are
also much larger than those of the spectral functions $C_{rc}(w,\Delta w)$ 
and $C_{cr}(w,\Delta w)$. The quantitative contribution of the two last 
spectral functions to the variance $\sigma_t^2(w,\Delta w)$ is therefore
negligible.
 
As shown in Refs.\ \cite{Wil87,Eck92}, the leading order contribution to the 
spectral function $C_{cc}(w,\Delta w)$ is proportional to the Fourier 
transform of the classical autocorrelation function of the Weyl transform of
the perturbing operator. By virtue of the ergodic theorem, the microcanonical 
average over the ergodic part of the energy shell appearing in the leading
order contribution can be replaced with a time (i.e. scaled action) average 
along any ergodic trajectory exploring this part in a uniform way. The 
resulting expression of the leading order contribution to the local variance 
$\sigma_c^2(w,\Delta w)$ is \cite{Boo95}
\begin{equation}
   \sigma_c^2(w,\Delta w)
 = {1\over\pi\rho_{0,c}(w)} {\rm Re} \int_0^\infty
   d\tilde s \, e^{i(\Delta w+i\epsilon) {\tilde s}} \, C_{{\tilde A},c}
   (\tilde s) \; .
\label{Semi}
\end{equation}
Here $C_{{\tilde A},c}(\tilde s)$ is the classical autocorrelation function 
of the Weyl transform ${\tilde A}(\tilde s)$, as computed along an ergodic  
trajectory of arbitrary large scaled action $S$, i.e.,
\begin{eqnarray}
 && C_{{\tilde A},c}(\tilde s) = \lim_{S\to\infty}{1\over S}\int_0^S d\tilde s'
    \nonumber \\  &\times&
   \left(\tilde A(\tilde s'+\tilde s/2)-\langle\hat A\rangle_c\right)
   \left(\tilde A(\tilde s'-\tilde s/2)-\langle\hat A\rangle_c\right) \; .
\label{Auto}
\end{eqnarray}
The ergodic theorem ensures that the limit on the r.h.s.\ of this equation is 
well defined and unique. The appearance in Eq.\ (\ref{Semi}) of a damping 
factor containing the width $\epsilon$ is a consequence of the smoothing of  
the $\delta$-function which is associated to differences in scaled energies 
in Eq.\ (\ref{Coirreg}). The formula in Eq.\ (\ref{Semi}) is well suited for 
the numerical computation of the classical contribution to the local variance 
$\sigma_c^2(w,\Delta w)$. It shows that, at the level of the leading order
contribution, the rescaled variance $\sigma_c^2(w,\Delta w) \rho_{0,c}(w)$ is a
function which depends only on the scaled energy difference $\Delta w$ and 
no more on the scaled energy $w$. 

\begin{figure}[b]
\vspace{5.2cm}
\includegraphics{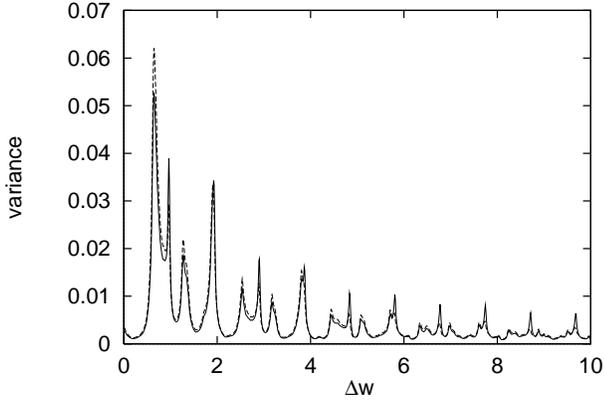}
\caption{\label{fig6} 
Rescaled variances of the distributions of non-diagonal transition matrix 
elements as a function of the scaled energy difference $\Delta w$ in the
case of the scaled energy $\tilde E=-0.2$.
Solid line: $\sigma_t^2(w,\Delta w)\rho_{0,t}(w)$; Dashed  
line: $\sigma_c^2(w,\Delta w)\rho_{0,c}(w)$.} 
\end{figure}

\begin{figure}[b]
\vspace{11.0cm}
\includegraphics{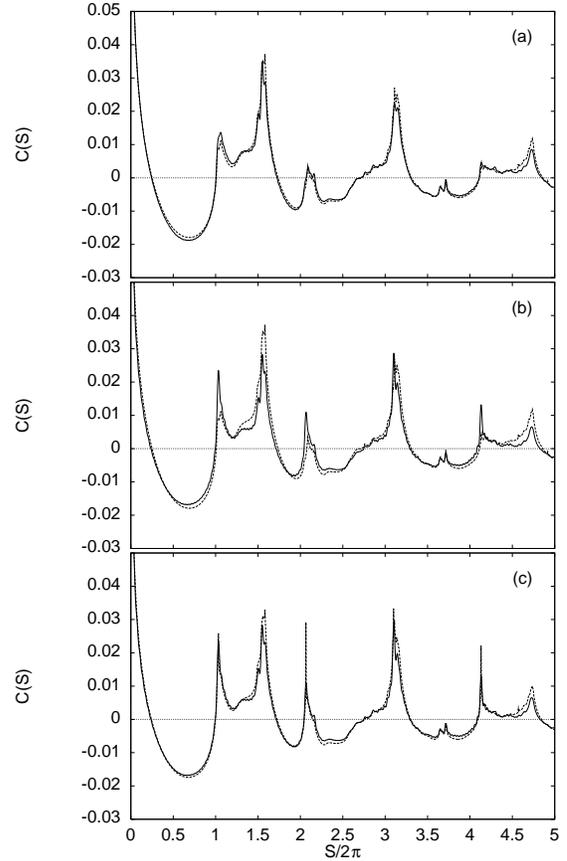}
\caption{\label{fig7} 
Fourier transform of the rescaled variances of the distributions of non-
diagonal transition matrix elements as a function of the scaled action 
${S/ 2 \pi}$ in the case of the scaled energy $\tilde E = -0.2$.
(a) Solid line: Fourier transform of $\sigma_c^2(w,\Delta w)\rho_{0,c}(w)$;
    Dashed line: Autocorrelation function $C_{{\tilde A},c}$.
(b) Solid line: Fourier transform of $\sigma_t^2(w,\Delta w)\rho_{0,t}(w)$;
    Dashed line: Autocorrelation function $C_{{\tilde A},c}$.
(c) Solid line: Fourier transform of $\sigma_t^2(w,\Delta w)\rho_{0,t}(w)$;    
    Dashed line: Weighted sum of autocorrelation functions $C_{{\tilde A},c}$ 
    and $C_{{\tilde A},r}$.} 
\end{figure}

\begin{figure}[b]
\vspace{11.0cm}
\includegraphics{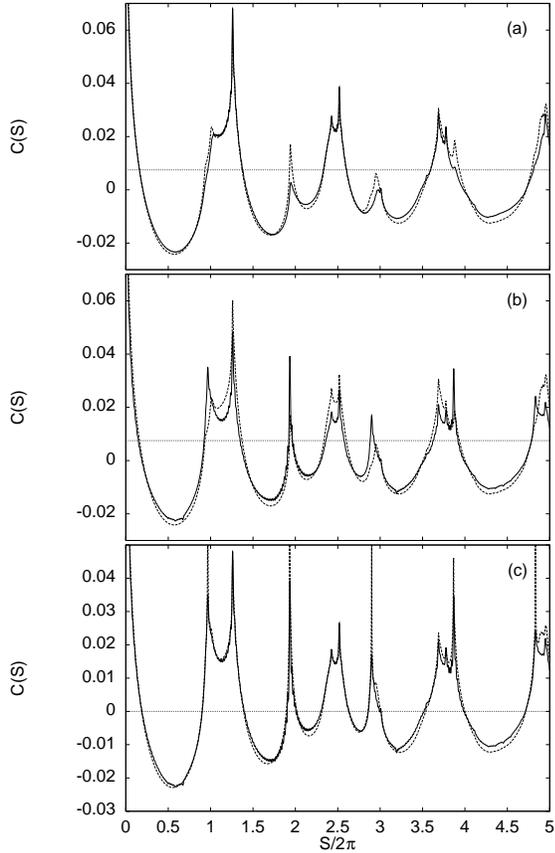}
\caption{\label{fig9} 
Same as Fig. 7 but in the case of the scaled energy $\tilde E=-0.316$.}
\end{figure}

As a first numerical study, it is interesting to compare the values of the 
rescaled variances $\sigma_t^2(w,\Delta w) \rho_{0,t}(w)$ and 
$\sigma_c^2(w,\Delta w) \rho_{0,c}(w)$ over a large range of values of the
scaled energy difference. This is done in Fig.\ \ref{fig6} for the case 
$\tilde E = -0.2$. The full (resp.\ dashed) curve corresponds to the first
(resp.\ second) rescaled variance. One sees that both curves are modulated by  
the period $S_r$ of the stable
\newpage
\phantom{}
\noindent
orbit lying in the plane perpendicular to the 
direction of the magnetic field since they exhibit peaks exactly at the 
integer multiples of the value $\Delta w=2\pi/S_r=0.968$. This modulation  
effect of the local variance was already pointed out in 
Refs.\ \cite{Boo95,Meh95}. One observes also that the two curves are only
slightly different from each other. This observation simply reflects the fact
that the number of transitions involving chaotic eigenstates is very much
larger than the number of transitions involving regular eigenstates in the
chosen example. Consequently, a
numerical analysis using the bare variances themselves would not allow to get  
precise enough values for the quantitative contribution to the total variance 
coming from the regular part of phase-space. As seen below, it is
necessary to use the Fourier transform of the rescaled variances in order to 
extract quantitative precise results from the 
numerical data. In practice, the full curve in Fig.\ \ref{fig6} has been 
calculated with the help of the numerical data built up from the distributions
of the non-diagonal matrix elements whereas the dashed curve has been
computed by using Eqs.\ \ref{Semi},\ \ref{Auto}. It is therefore interesting 
to compare the predictions of the expression of the leading order 
contribution to the local variance $\sigma_c^2(w,\Delta w)$ with the 
numerical values obtained from the distributions of non-diagonal matrix 
elements coupling chaotic eigenstates together. This is done in 
Fig.\ \ref{fig7}a for the case $\tilde E = -0.2$ and in Fig.\ \ref{fig9}a
for the case $\tilde E = -0.316$. In both figures, the full curve represents
the Fourier transform $C(S)$ of the rescaled variance 
$\sigma_c^2(w,\Delta w) \rho_{0,c}(w)$, i.e.,
\begin{equation}
 C(S) = e^{{\epsilon} S} \int_{-\infty}^{+\infty} 
        d(\Delta w) \cos(\Delta w S) \,
        \sigma_c^2(w,\Delta w) \, \rho_{0,c}(w)  \; ,
\label{Foucha}
\end{equation}
as calculated from the relevant numerical data. On the other hand, the dashed
curve represents the classical autocorrelation function $C_{{\tilde A},c}(S)$,
Eq.\ (\ref{Auto}), as computed 
\newpage
\phantom{}
\noindent
along an arbitrary ergodic trajectory. 
Eq.\ (\ref{Semi}) predicts that both curves are identical, i.e.\
$C(S) = C_{{\tilde A},c}(\tilde S)$ . Figs.\ \ref{fig7}a and \ \ref{fig9}a
show that they agree well with each other over the whole range of values of 
the scaled action $S$. The values of the local variance 
$\sigma_c^2(w,\Delta w)$ can therefore be reproduced with an excellent 
precision by the leading order contribution alone.

\begin{figure}[b]
\vspace{4.5cm}
\includegraphics{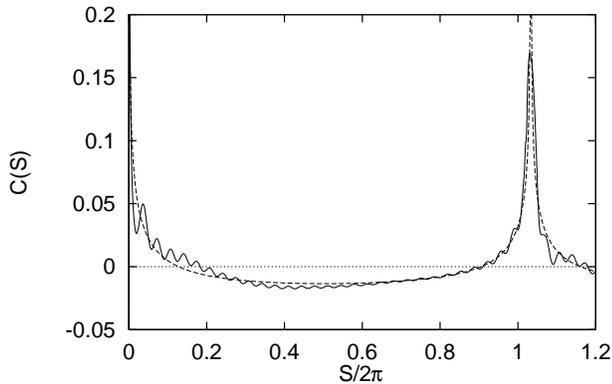}
\caption{\label{fig8} 
Autocorrelation function $C_{{\tilde A},r}$ as a function of the scaled 
action ${S/ 2 \pi}$ 
in the case of the scaled energy $\tilde E = -0.2$. Solid 
line: Computation with Eq.\ \ref{Ftre}; Dashed line: Computation with 
Eq.\ \ref{Corregu}.}
\end{figure}

The expression of the spectral function $C_{rr}(w,\Delta w)$ in the
semiclassical regime is obtained much in the same way as the expression of the
spectral function $D_r(w)$, Eq.\ (\ref{pois}). Indeed, the contribution of a
stable periodic orbit and all its repetitions to $C_{rr}(w)$ has an 
expression which is analogous to the one given in Eq.\ (\ref{misu}). The only
difference is that the amplitude $A_r$ is now replaced with the Fourier  
transform of the autocorrelation function of the Weyl transform of 
the perturbing operator along the stable periodic orbit
\cite{Wil87,Eck92,Meh95}. The formal resummation of all contributions leads 
to the semiclassical expression
\newpage
\phantom{}
\begin{equation}
 C_{rr}(w,\Delta w) = {\rho_{0,r}(w)\over\pi} {\rm Re} \int_0^\infty
   d\tilde s \, e^{i(\Delta w+i\epsilon) {\tilde s}} \, C_{{\tilde A},r}
   (\tilde s)  \; , 
\label{Regrega}
\end{equation}
with the autocorrelation function $C_{{\tilde A},r}(\tilde s)$ given by 
%
\begin{eqnarray}
 && C_{{\tilde A },r}(\tilde s) 
   = {1\over S_r} \oint d\tilde s'  \nonumber \\  &\times&
   \left(\tilde A(\tilde s'+\tilde s/2)-\langle\hat A\rangle_r\right) 
   \left(\tilde A(\tilde s'-\tilde s/2)-\langle\hat A\rangle_r\right)  \; .
\label{Corregu}
\end{eqnarray}
The autocorrelation function is a periodic function with the same period 
$S_r$ as the one of the orbit \cite {Wil87,Meh95}. In order to check the
validity of the semiclassical expression of $C_{rr}(w,\Delta w)$, it is 
easier to take the Fourier transform with respect to the scaled energy 
difference $\Delta w$ on both sides of Eq.\ (\ref{Regrega}) and to integrate
subsequently with respect to the scaled energy $w$. If $N$ is the number of
regular states in the used spectrum, one gets the following expression of the 
autocorrelation function 
\begin{eqnarray}
  C_{{\tilde A},r}(S) &=& {1\over N} \sum_{m_r,n_r} 
  |\langle m_r|\hat A - {\langle {\hat A}\rangle_r}|n_r\rangle|^2 \nonumber \\
 &\times&   \cos\left(\left(w_m^{(r)}-w_n^{(r)}\right)S\right)  \; .
\label{Ftre}
\end{eqnarray}
Fig.\ \ref{fig8} compares the values of the autocorrelation function, as 
computed with the help of Eq.\ (\ref{Ftre}) (full curve) and 
Eq.\ (\ref{Corregu}) (dashed curve). 
The comparison is done in the case $\tilde E = -0.2$ 
($N = 120$) and over one period of the stable trajectory. It is seen that the 
agreement between both curves is good, except in the vicinity of the values
$S = 0$ and $S = S_r$ for which the expression in Eq.\ (\ref{Corregu}) is 
singular. The full curve would reproduce this singular behavior all the 
better as the number of regular states used in the computation would be larger.
The finite number of regular states is also responsible for the observed 
little 
discrepancies between both curves in the considered interval of values of the 
scaled action. In spite of these discrepancies, the general good agreement 
between the full and the dashed curve allows to conclude that the semiclassical
expression given in Eq.\ (\ref{Regrega}) can be used to calculate the values 
of the spectral function $C_{rr}(w,\Delta w)$. By analogy with 
Eq.\ (\ref{Varcha}), one can also introduce a local variance 
$\sigma_r^2(w,\Delta w)$ corresponding to the distributions of 
non-diagonal matrix elements coupling regular eigenstates together, defined as 
\begin{equation}
           \sigma_r^2(w,\Delta w) 
           = {C_{rr}(w,\Delta w)\over \left(\rho_{0,r}(w)\right)^2}  \; .
\label{Vregu}
\end{equation}
It is to be noted that the spectral functions which are associated to the
distributions of non-diagonal matrix elements involving chaotic 
(Eqs.\ \ref{Semi},\ \ref{Auto}) and regular 
(Eqs.\ \ref{Regrega},\ \ref{Corregu}) states have similar classical  
expressions. As in the case of the diagonal matrix elements, each expression 
takes into account the classical trajectory which is related to the 
component of phase-space connected to the studied subset of non-diagonal 
matrix elements. 

\begin{figure}[b]
\vspace{7.2cm}
\includegraphics{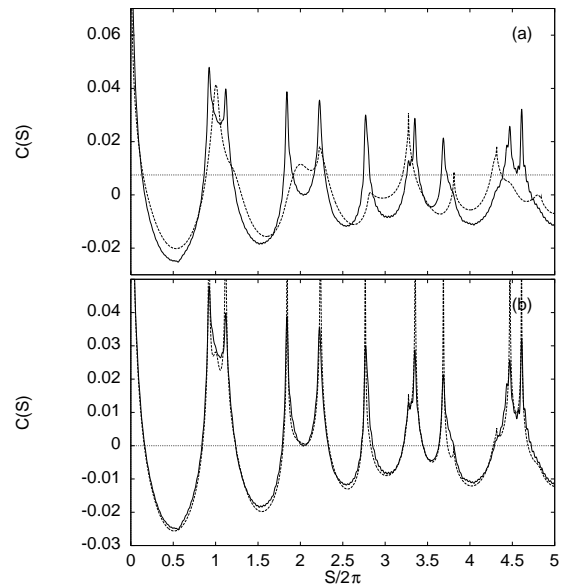}
\caption{\label{fig10}
Fourier transform of the rescaled variance 
$\sigma_t^2(w,\Delta w)\rho_{0,t}(w)$ of the distributions of non-diagonal
transition matrix elements as a function of the scaled action ${S/2 \pi}$ 
in the case of the scaled energy $\tilde E = -0.4$.
(a) Solid line: Fourier transform of $\sigma_t^2(w,\Delta w)\rho_{0,t}(w)$;
    Dashed line: Autocorrelation function $C_{{\tilde A},c}$.
(b) Solid line: Fourier transform of $\sigma_t^2(w,\Delta w)\rho_{0,t}(w)$;    
    Dashed line: Weighted sum of autocorrelation functions $C_{{\tilde A},c}$ 
    and $C_{{\tilde A},r}$.} 
\end{figure}

It has been checked previously that a very close estimate of the values of the
total spectral function $C(w,\Delta w)$ is obtained through the sum of the two 
spectral functions $C_{cc}(w,\Delta w)$ and $C_{rr}(w,\Delta w)$. Consequently,
by virtue of Eqs.\ \ref{Var},\ \ref{Varcha} and \ \ref{Vregu}, the rescaled
variance which is associated to the distributions of all non-diagonal
matrix elements can be written as
\begin{eqnarray}
 \sigma_t^2(w,\Delta w){\rho_{0,t}(w)}
   &=& \left({\rho_{0,c}(w)\over \rho_{0,t}(w)}\right)
       \sigma_c^2(w,\Delta w)\rho_{0,c}(w) \nonumber\\
   &+& \left({\rho_{0,r}(w)\over \rho_{0,t}(w)}\right)
       \sigma_r^2(w,\Delta w)\rho_{0,r}(w)  \; .
\label{Tovar}
\end{eqnarray}
As in the case of the mean value $\langle \hat A\rangle_t$, 
Eq.\ (\ref{centro}), the contribution to the rescaled variance 
$\sigma_t^2(w,\Delta w){\rho_{0,t}(w)}$ of the ergodic (resp.\ regular)
component of phase-space is weighted by the ratio of the mean density of states
of the subset of chaotic (resp.\ regular) states to the mean total density of 
states. Equivalently, taking the Fourier transform on both sides of 
Eq.\ (\ref{Tovar}) with respect to the scaled energy difference $\Delta w$, one
obtains the following formula with the help of Eqs.\ \ref{Semi},\ \ref{Regrega}
and \ \ref{Vregu}  
\begin{eqnarray}
  && e^{{\epsilon} S} \int_{-\infty}^{+\infty} d(\Delta w) \cos(\Delta w S) \, 
     \sigma_t^2(w,\Delta w) \, \rho_{0,t}(w) \nonumber \\ 
  &=& {\left({\rho_{0,c}(w)\over \rho_{0,t}(w)}\right)}C_{{\tilde A},c}(S)
    +{\left({\rho_{0,r}(w)\over \rho_{0,t}(w)}\right)}C_{{\tilde A},r}(S) \, .
\label{Ovar}
\end{eqnarray}
\newpage
\phantom{}
\noindent
The numerical comparison between both sides of this equation is done in 
Figs.\ \ref{fig7}b,c for the case $\tilde E = -0.2$. In Fig.\ \ref{fig7}b, 
the full curve represents the 
Fourier transform of the rescaled variance whereas the dashed curve represents
the autocorrelation function $C_{{\tilde A},c}(S)$ alone. One sees that both
curves do not agree so much with each other. One has to compare 
Fig.\ \ref{fig7}b with Fig.\ \ref{fig7}c in order to appreciate the 
quantitative improvement which is brought in the weighted sum by the 
contribution originating from the stable periodic orbit. In this last 
figure, the full curve represents again the Fourier transform of the rescaled 
variance whereas the dashed curve represents now the weighted sum of the 
autocorrelation functions on the r.h.s.\ of Eq.\ (\ref{Ovar}). Contrary to 
Fig.\ \ref{fig7}b, both curves agree now well with each other over the whole 
range of values of $S$. The improvement is especially noticeable in the 
immediate vicinity of the peaked structures located at positions which are
multiple integers of $S_r/2\pi=1.03305$. This is due to the fact that the 
autocorrelation function $C_{{\tilde A},r}(S)$ contributes mostly to the 
Fourier transform in the immediate vicinity of these positions, as shown by
Fig.\ \ref{fig8}. Fig.\ \ref{fig7}c provides also another confirmation of the 
fact that the spectral functions $C_{cr}(w, \Delta w)$ and 
$C_{rc}(w, \Delta w)$ give really a negligible quantitative contribution to the
local variance. One can therefore conclude that Eq.\ (\ref{Tovar}) is able to 
reproduce the values of the local variance with an excellent precision. This 
equation generalizes to the case of a scaled system with a mixed phase-space 
the expression of the local variance given in Refs. \cite{Wil87,Eck92} for a 
scaled system with an ergodic phase-space.   
As in the case of the diagonal matrix elements, the generalization of 
Eq.\ (\ref{Tovar}) to a situation with several regular components in
phase-space is straightforward. Indeed, one has to add as many weighted 
rescaled variances $\sigma_r^2(w,\Delta w){\rho_{0,r}(w)}$ as there are stable 
periodic orbits in the underlying classical dynamics, as illustrated by the
remaining figures. On the one hand, Fig.\ \ref{fig9}b 
(resp.\ Fig.\ \ref{fig10}a) compares the Fourier transform of the rescaled 
variance $\sigma_t^2(w,\Delta w){\rho_{0,t}(w)}$ (full curve) with the 
autocorrelation function $C_{{\tilde A},c}(S)$ (dashed curve) in the case 
\newpage
\phantom{}
\noindent
$\tilde E = -0.316$ (resp.\ $\tilde E = -0.4$). As in the previous case, there 
are many discrepancies between both curves because the contributions coming 
from the stable periodic orbits are missing. On the other hand, 
Fig.\ \ref{fig9}c (resp.\ Fig.\ \ref{fig10}b) compares the Fourier transform 
of the rescaled variance (full curve) with the weighted sum of autocorrelation
functions (dashed curve) in the case $\tilde E = -0.316$ 
(resp.\ $\tilde E = -0.4$). The sum now contains the properly weighted 
contributions originating from the stable periodic orbits which have been 
identified in the previous Section. In both cases, the very good agreement 
between the two curves confirms the validity of the generalization of 
Eq.\ (\ref{Tovar}).

\section{Conclusion}
This paper has been devoted to the study of the quantitative contributions  
of the different components making up a mixed phase-space to the value of the 
mean characterizing the distribution of diagonal transition matrix elements
and to the value of the variance characterizing the distributions of 
non-diagonal transition matrix elements. With the help of numerical 
computations done in the framework of the Hydrogen atom in a strong magnetic
field, it has been shown that these contributions can be well identified 
in the semiclassical regime. The computations have confirmed that the 
leading order contribution of the ergodic component to the mean is equal to 
the average of the Weyl transform $\tilde A$ of the perturbing operator 
$\hat A$ along an arbitrary ergodic trajectory. They have also confirmed that 
the leading order contribution of the same component to the variance is
proportional to the Fourier transform of the autocorrelation function of
$\tilde A$, this autocorrelation function being also computed along an 
arbitrary ergodic trajectory. On the other hand, it has been found that the
contribution of each regular component to the mean is equal to the average
of $\tilde A$ around the corresponding stable orbit. It has also been found
that the contribution of each such component to the variance is proportional to
the Fourier transform of the autocorrelation function of $\tilde A$, this 
autocorrelation function being again computed around the corresponding stable 
orbit. For each studied quantity, the contributions coming from the ergodic 
and regular components have therefore similar expressions, each expression 
taking into account the particular classical trajectory which is related to 
the considered component. The stable periodic orbits provide a convenient 
method to compute the contributions of the various regular components to mean 
and variance with high accuracy. This method is different from the one which
has been proposed by Robnik and Prosen for the same purpose. As a final step,
it has been shown that mean and variance can be expressed as a weighted sum of
the contributions of all different components belonging to the mixed 
phase-space. The weight appearing in front of a given contribution has been  
identified as the ratio of the mean density of states of the corresponding
component to the mean total density of states of the system. Although the 
study has been done for a particular scaling system, the results presented in 
this paper are relevant to all generic scaling systems with a small number of 
degrees of freedom having a mixed phase-space. 

\acknowledgments
\vskip -4pt
We thank B.\ Mehlig and K.\ M\"uller for discussions and a referee 
for useful comments. This work was supported in part by the 
Deutsche For\-schungsgemeinschaft (Sonder\-for\-schungs\-be\-reich No.\ 237).


\begin{references}
\bibitem{Boo96:Erratum}
  Eq. (1) of Ref.\ \cite{Boo96} does not give the accurate expression of the
  microcanonical average. One obtains the accurate expression by replacing
  the scaling parameter (which is called $z$ in Ref.\ \cite{Boo96}) by the
  scaled energy $\tilde E$ in the numerator and denominator of Eq. (1).
\end{references}
\end{document}